\documentclass[aip,rsi, reprint]{revtex4-1}
\usepackage{graphicx}
\usepackage{dcolumn}
\usepackage{bm}
\usepackage{subcaption}

\begin{document}
\title{A Low-Frequency Torsion Pendulum with Interferometric Readout}

\author{M.P. Ross}
\email[]{mpross2@uw.edu}
\author{K. Venkateswara}
\author{C.A. Hagedorn}
\affiliation{Center for Experimental Nuclear Physics and Astrophysics, University of Washington, Seattle, Washington,
98195, USA}
\author{C.J. Leupold}
\affiliation{University of Washington Bothell, Bothell, WA 98011, USA}
\author{P.W.F. Forsyth}
\affiliation{OzGrav-ANU, Centre for Gravitational Astrophysics, College of Science, The Australian National University, Acton, ACT 2601, Australia}
\author{J.D. Wegner}
\author{E.A. Shaw}
\author{J.G. Lee}
\author{J. H. Gundlach}
\affiliation{Center for Experimental Nuclear Physics and Astrophysics, University of Washington, Seattle, Washington,
98195, USA}

\begin{abstract}
We describe a torsion pendulum with a large mass-quadrupole moment and a resonant frequency of 2.8~mHz, whose angle is measured using a Michelson interferometer. The system achieved noise levels of $\sim~200\ \text{prad}/\sqrt{\text{Hz}}$ between 0.2-30 Hz and $\sim10\ \text{prad}/\sqrt{\text{Hz}}$ above 100 Hz. Such a system can be applied to a broad range of fields from the study of rotational seismic motion and elastogravity signals to gravitational wave observation and tests of gravity.
\end{abstract}
\maketitle

\section{Introduction}
\quad Since the days of Coulomb\cite{coulomb}, Cavendish\cite{cavendish}, and E\"otv\"os \cite{eotvos}, torsion balances have been used for a wide variety of precision measurements. Over the years they have allowed precise measurement of the gravitational constant \cite{bigG}, tests of the behavior of gravity \cite{shortRange, EP}, hunts for dark matter \cite{uldm, spinDM}, and searches for novel fifth forces \cite{spin}. Torsion balances used to search for novel interactions are typically designed to minimize gravitational couplings while maximizing torques caused by the proposed novel interaction. This is done by minimizing the mass-multipole moments of the pendulum while maintaining a dipole for the charge corresponding to the novel interaction. Additionally, the apparatus developed for these searches take great care to shield any spurious couplings between the pendulum and the environment. These together tend to yield highly-symmetric pendulums with relatively small moments of inertia.

However, recently large-moment torsion balances have begun to be built with the goal of observing gravitational waves\cite{toba} and measuring gravity gradients\cite{torpedo}. These balances are designed to maximize gravitational couplings while minimizing the torsional resonant frequency. To precisely measure minute gravitational effect, these systems employ sensitive interferometric angular readouts and precision control schemes. Additionally, similar apparatus have been developed to develop technology for future gravitational wave observatories\cite{ciani} and as novel inertial-isolation schemes\cite{sixD}. This class of instruments has also been proposed to study elastogravity signals from seismic events.\cite{elasto}

In this paper, we describe a prototype large-moment torsion balance, the Michelson Interferometer Torsion-balance (MINT), that employs a Michelson interferometer to achieve sub-$\text{nrad}/\sqrt{\text{Hz}}$ angular readout of a low resonant frequency, dumbbell-shaped pendulum.

\section{Mechanical Design}

\begin {figure}[!h]
\centering
\includegraphics[width=0.5\textwidth]{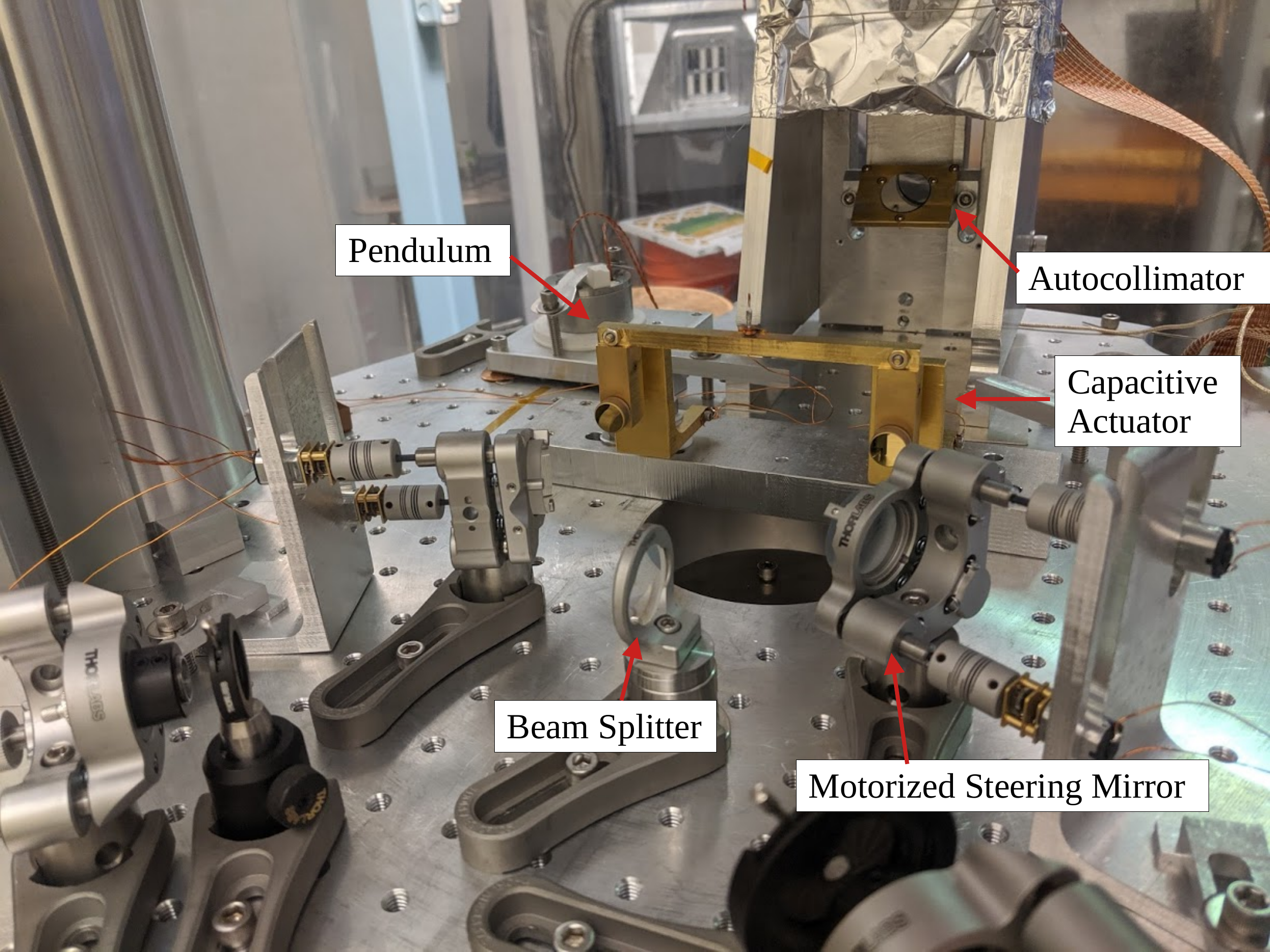}
\caption{A photograph of the instrument showing the pendulum, interferometer optics, capacitive actuators, and autocollimator optics. Not shown is the magnetic damper and alignment stage.}
\label{pic}
\end{figure}

\begingroup
\setlength{\tabcolsep}{10pt} 
\renewcommand{\arraystretch}{1.5} 
\begin{table}[!h]
\centering
\begin{tabular}{|c|c| }
\hline
Parameter & Value \\ 
\hline
$\kappa$ & $2.89(2) \times 10^{-8} \textrm{ N m/rad}$\\
$I$ & $9.40(5) \times 10^{-5} \textrm{ kg m}^2$\\ 
$f_0$ & 2.79(4) mHz\\
$Q$ & $3.0(5)\times 10^3$\\
\hline
\end{tabular}
\caption{Parameters describing the torsional mode of the MINT balance where $\kappa$ is the torsional spring constant, $I$ is the mass-moment of inertia, $f_0$ is the resonant frequency, and $Q$ is the observed quality factor.}
\label{table}
\end{table}

\endgroup

\quad The MINT balance, shown in Figure \ref{pic}, consists of a 11.4-cm-wide aluminum pendulum hung from a 20-$\mu$m-diameter and 8.9-cm-long tungsten fiber. The width of the pendulum allowed for increased angular sensitivity and a lower resonant frequency of the balance via an increased lever-arm and a larger moment of inertia, respectively.

The pendulum is suspended from an intermediate copper mass which itself is suspended from a rigid structure with a 75-$\mu$m-thick 7.6-cm-long ``pre-hanger'' tungsten fiber and a beryllium-copper leaf spring. This intermediate mass is place within an exterior magnetic field to form an eddy current damper which is detailed in Section \ref{damper}. This damper is attached to a translation stage which can move in both horizontal directions as well as rotate about vertical. Additionally, two plane electrodes are placed behind the pendulum to allow for active control.

\begin {figure}[!h]
\centering
\includegraphics[width=0.5\textwidth]{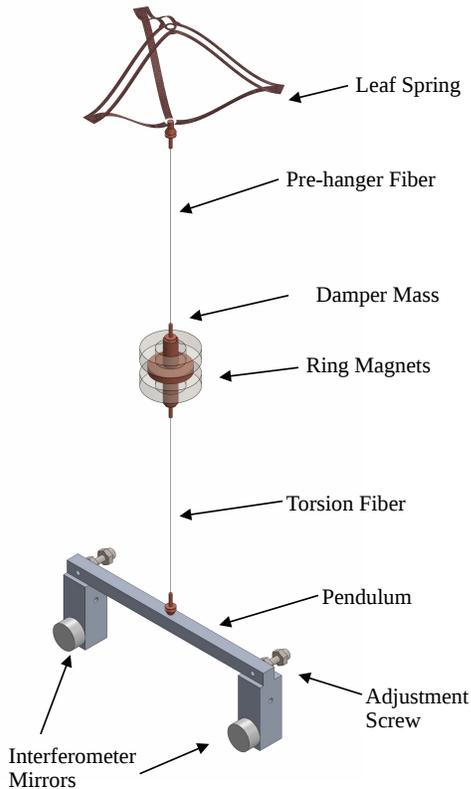}
\caption{CAD rendering of the MINT pendulum and eddy current damper assembly.}
\label{Model}
\end{figure}

A CAD rendering of the pendulum is shown in Figure~\ref{Model}. To avoid spurious forces and to decrease losses due to gas damping, the pendulum is housed in a vacuum chamber held at $\sim$ 0.1 mPa by continuous pumping with a turbo pump. The parameters describing the torsional mode of the MINT balance are shown in Table~\ref{table}.

\section{Eddy Current Damper} \label{damper}

\quad In order to simplify the system, one would prefer to operate an instrument with a single degree of freedom. However, any torsion fiber will also permit ``swing'' modes whose restoring force is predominantly gravity and not the elasticity of the fiber. The swing modes are readily excited due to ambient seismic motion and can become the highest amplitude modes if not controlled appropriately. 

In the MINT apparatus, these modes are passively damped by suspending the pendulum from an intermediate stage consisting of a copper disk suspended by both a vertical spring and a thin fiber, shown in Figure \ref{Model}. The vertical spring is formed by a laser cut beryllium copper leaf spring which gives a vertical, ``bounce'', mode of $\sim$~3~Hz. Above this frequency, the balance remains vertically inertial thereby isolating it from high frequency vertical seismic motion.


The damper mass is placed within the magnetic field of two ring magnets which are held by a separate rigid mount. When the pendulum swings, the motion of the disk within the non-uniform magnetic field drives eddy currents which damp the motion. Torsional motion, on the other hand, does not drive significant eddy currents due to the cylindrical symmetry of the magnets and the increased stiffness of the pre-hanger fiber. Thus, this system allows torsional motion with low-loss but provides high-loss for swing and vertical motion.

\section{Optical Readout} \label{readout}

\quad Two independent, in-vacuum optical readouts are operated to allow for both coarse and fine angular measurements of the pendulum. A two-dimensional autocollimator senses the angle of a mirror attached to the center of the pendulum with a sensitivity of $\sim20\ \text{nrad}/\sqrt{\text{Hz}}$. The autocollimator was used for initial alignment and cross-calibration. 

\begin {figure}[!h]
\centering
\includegraphics[width=0.5\textwidth]{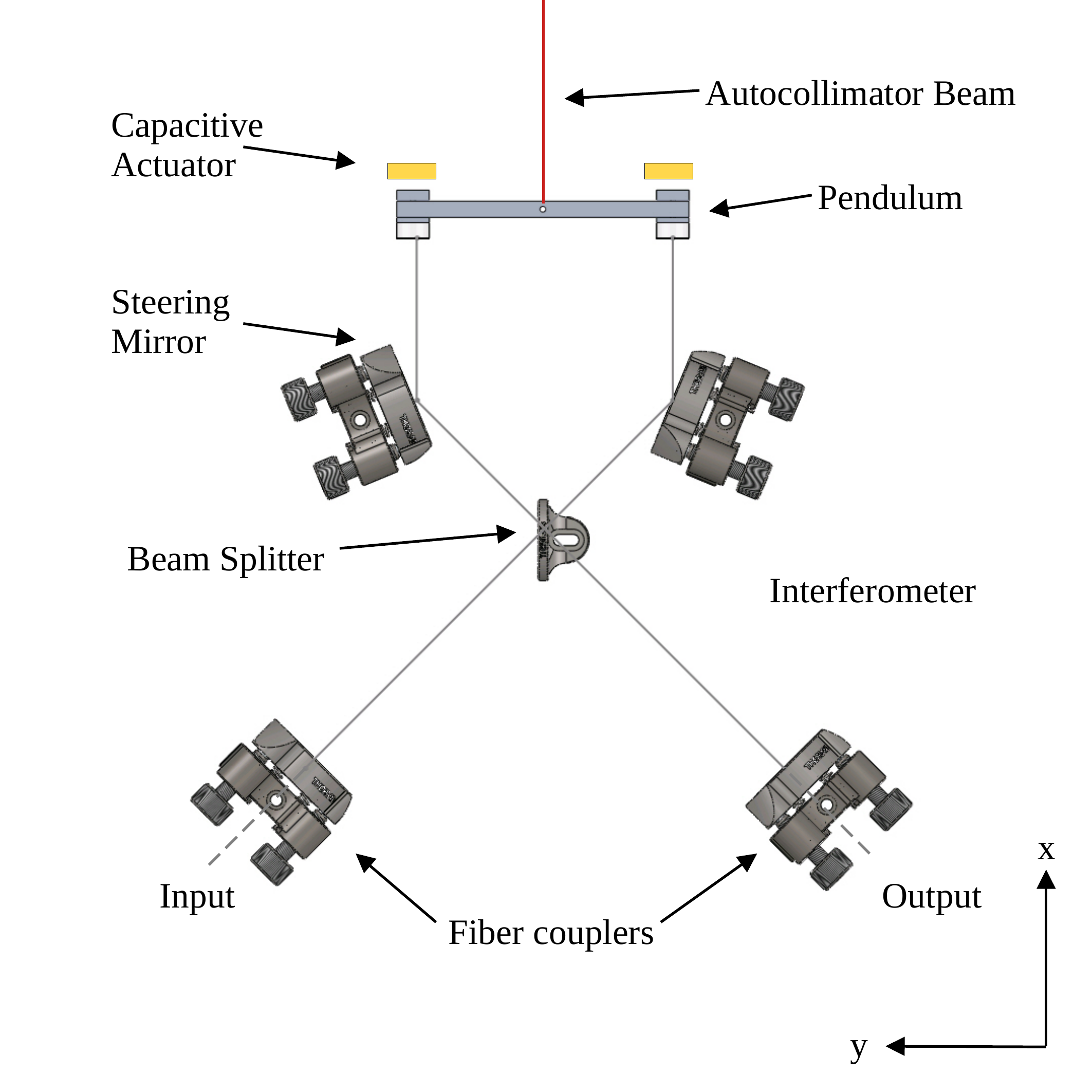}
\caption{Optical layout of the MINT apparatus. The autocollimator is comprised of the beam path at the top of the image while the Michelson interferometer is formed by the bottom beam paths.}
\label{optics}
\end{figure}

The primary readout was formed by a Michelson interferometer, shown in Figure \ref{optics}, whose arms were formed by two mirrors attached to opposite ends of the pendulum. Angular motion induces a phase difference, $\phi$, between the two arms given by:
\begin{equation}
\phi = 4 \pi \frac{r}{\lambda} \theta
\end{equation}
where $r$ is the distance from the fiber to the location of the beam on the mirror, $\lambda$ is the wavelength of light, and $\theta$ is the angle of the pendulum. This phase difference causes changes in the light intensity entering the output port. A 1310-nm fiber coupled laser was connected to the input port of the interferometer and the output port was attached to a in-air photodiode. Both of these connections were made via a pair of teflon optical-fiber vacuum feedthroughs \cite{FiberCouple}.

\section{Controls}

In order to operate the interferometer in a linear regime, the pendulum is locked in feedback using two parallel-plate capacitive actuators. The lock is achieved in two stages, first the autocollimator is used as the feedback sensor to decrease the amplitude of the torsional mode and provide a course alignment. Then the feedback control is switched to the interferometer readout to allow for low noise operation. The feedback employs a PID loop whose unity gain frequency was in the 10-30 mHz range. The physical voltage-to-force gain of the capacitive actuator depends on the gap between the actuator and the pendulum. A relatively large gap ($\sim$ 1-cm) was chosen to minimize control noise while providing sufficient low-frequency control. This design decreases the influence of actuation noise in the frequency range of interest while allowing for control of the torsional resonance.

\section{Cross Coupling Minimization}\label{cross}

Any coupling of non-torsional motion to the angular readout decreases the performance of the measurements and thus must be minimized. Additionally, although the magnetic damper described in Section \ref{damper} decreases the motion in the swing modes, the residual motion in the swing resonances adds significant artifacts in the readout if these couplings aren't minimized. In this section, we give a simple description of the methods used to decrease the coupling of the readout to these swing modes. For a more detailed study of such couplings see \citet{cross-couple}.

\begin{widetext}

\begin{centering}
\begin {figure}[!h]
\includegraphics[width=0.9\textwidth]{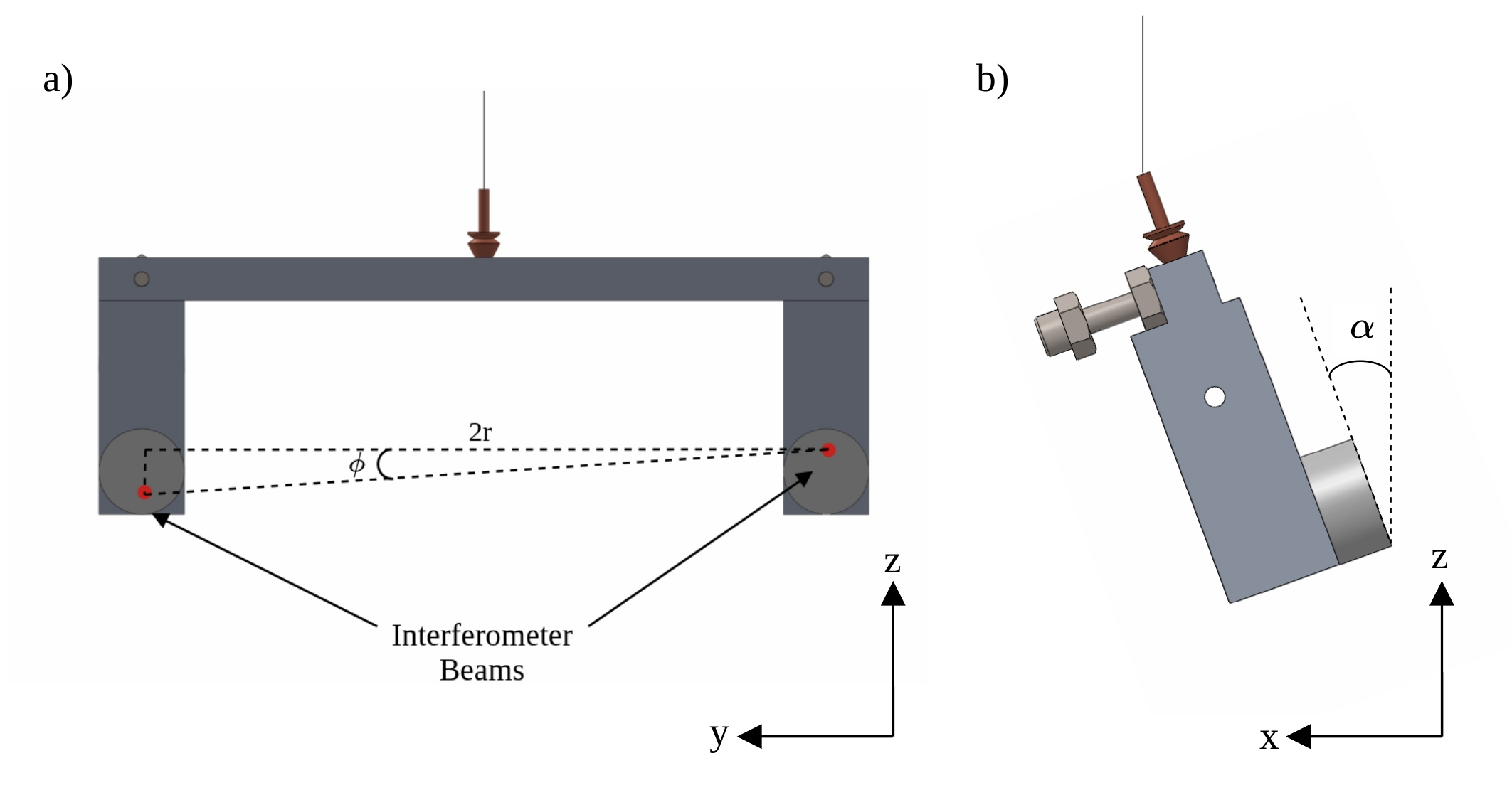}
\caption{Geometry of cross-coupling mechanisms discussed in Section \ref{cross} with misalignments exaggerated for clarity.}\label{geo}
\end{figure}
\end{centering}
\pagebreak
\end{widetext}

If the interferometer beams hit at different heights on the mirrors of the pendulum, as shown in Figure \ref{geo}a, then the swing motion about the y-axis couples to the interferometer readout. This is due to the two arms effectively sensing pendulums of different lengths. The path length difference due to this follows:
\begin{equation}
\Delta \delta\ =\ 2\ r\ \tan\phi\ \sin\theta_y
\end{equation}
where $\Delta \delta$ is the difference in path length, $r$ is the distance from the center of the pendulum to the center of the mirrors, $\phi$ is the angle between the beam spots, and $\theta_y$ is the angle of the pendulum about the y-axis with respect to vertical. This coupling was minimized by tilting the optical table about the x-axis, which rotates the beam spots while keeping the angle of the fiber fixed relative to the local vertical. This angle was changed to minimize the observed motion at the y-axis swing resonant frequency.

Additionally, if the faces of the mirrors are not parallel to the fiber, as shown in Figure \ref{geo}b, the swing motion about the x-axis couples due to the change in angle of the faces of the mirrors. Assuming that both mirrors are inclined by the same amount, this coupling follows:
\begin{equation}
\Delta \delta=2\ L\ \tan\alpha\ \sin\theta_x
\end{equation}
where $L$ is the length of the pendulum, $\alpha$ is the angle of the mirrors with respect to vertical, and $\theta_x$ is the angle of the pendulum about the x-axis with respect to vertical. This coupling was minimized by iteratively shifting trim screws placed at the back of the pendulum. By shifting the screws the center of mass of the pendulum shifted in the x-direction which caused the pendulum to tip about the y-axis.

With these two methods, the coupling of the interferometer to swing motion was decreased to $\sim$~$10^{-4}$~rad/rad. The residual effect of the swing resonance was comparable to the observed noise floor. In addition to decreasing the noise at the swing resonant frequency, this cross-coupling minimization decreased broadband noise due to horizontal tilts of the instrument driven by ambient seismic motion.

\section{Noise Performance}

\begin {figure}[!h]
\centering
\includegraphics[width=0.5\textwidth]{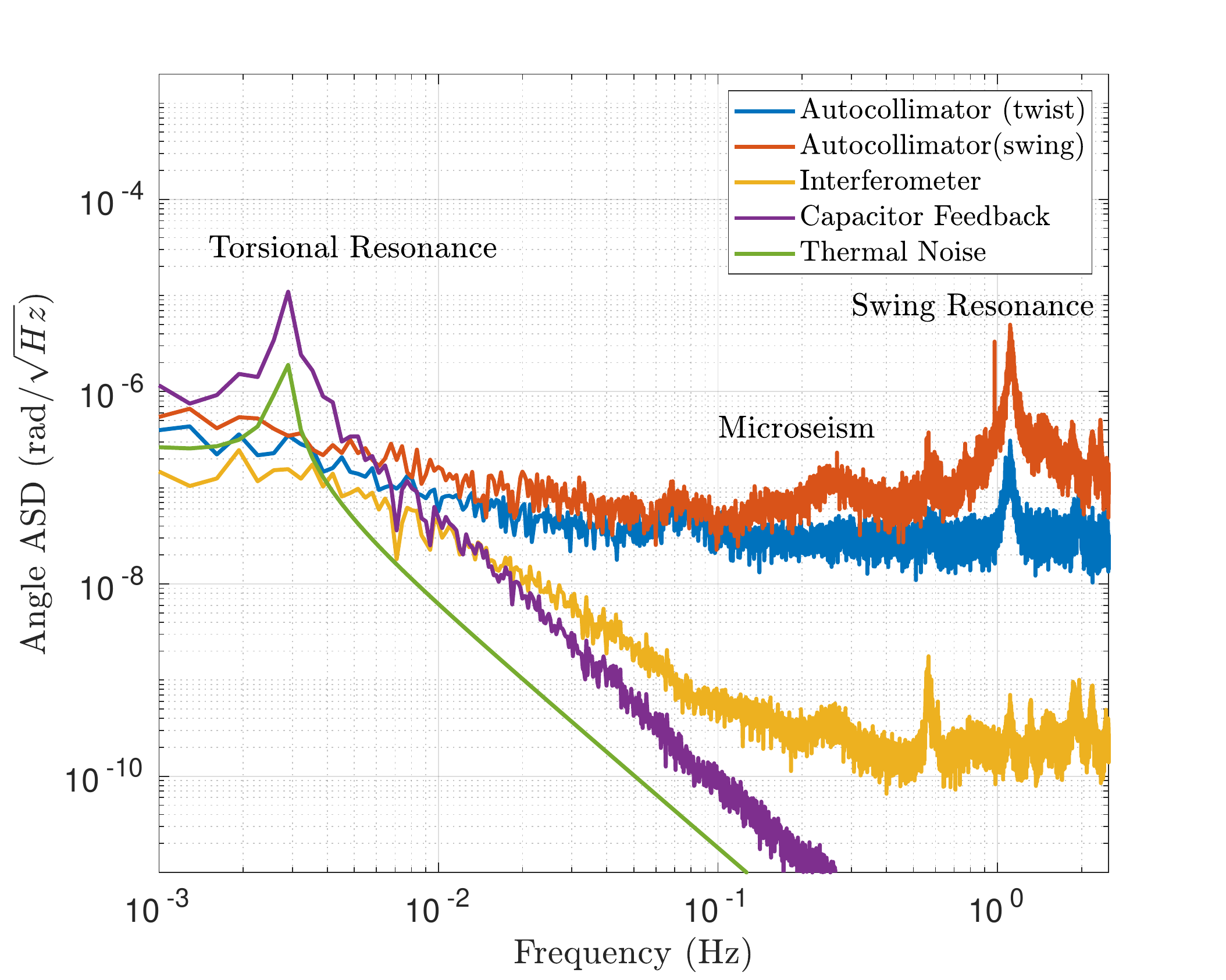}
\caption{Amplitude spectral density of the angle noise of both autocollimator directions, the interferometer readout, and the capacitor feedback. Also shown is the expected thermal noise limit of the pendulum.}
\label{ASD}
\end{figure}

With the cross coupling to non-torsional modes minimized, the instrument achieves the noise performance shown in Figure~\ref{ASD}. Since the pendulum is in active feedback, the readout of the instrument is the sum of the angular equivalent of capacitor feedback signal and the output of the interferometer. The feedback voltage was calibrated into angle units by injecting a small offset voltage onto the capacitors and measuring the interferometer response. Also shown is the motion measured by both axes of the autocollimator. Here we assume the absence of a signal therefore we treat all measured angles as noise.

\begin {figure}[!h]
\centering
\includegraphics[width=0.5\textwidth]{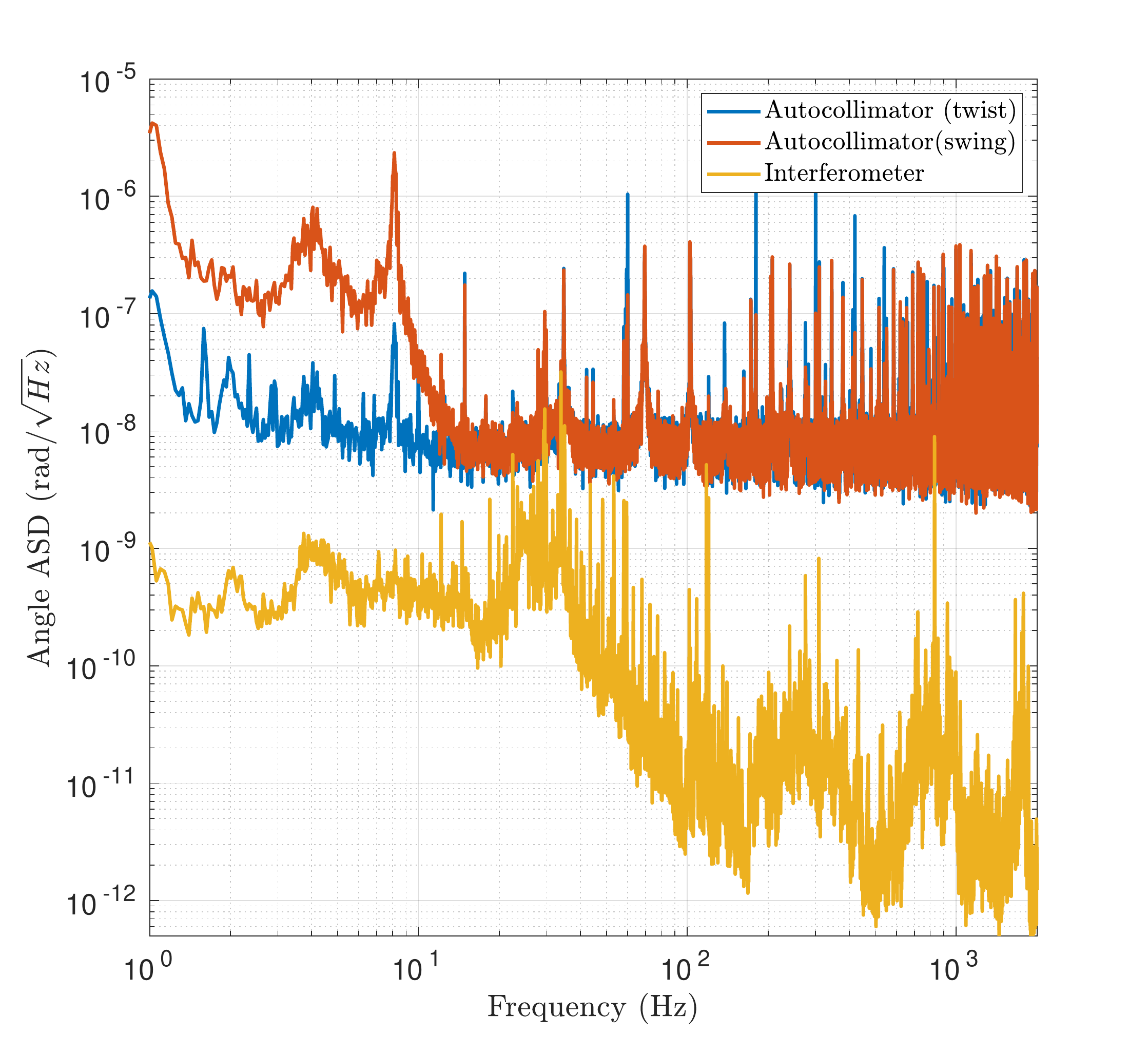}
\caption{High frequency amplitude spectral density of the angle noise of both autocollimator directions and the interferometer readout.}
\label{ASD_HF}
\end{figure}

The high frequency performance is shown in Figure~\ref{ASD_HF}. At these frequencies, the feedback contribution to the readout is negligible and is thus omitted. These spectra were taken at separate but subsequent times due to file size constraints.

It is apparent that the instrument has a noise floor of $\sim200\ \text{prad}/\sqrt{\text{Hz}}$ between 0.2-30 Hz and $\sim10\ \text{prad}/\sqrt{\text{Hz}}$ above 100 Hz. Below 0.2 Hz, the noise rises significantly to $\sim10\ \text{nrad}/\sqrt{\text{Hz}}$ at 30 mHz. This rise is believed to be due to both residual seismic noise coupling and temperature effects. In later runs, the noise in the 0.5-30 mHz range was reduced to within a factor of $\sim2$ of the suspension thermal noise by surrounding the pendulum with a thin aluminum housing. However, during these runs the $>$100 mHz noise was contaminated by increased seismic activity.

Multiple structures are apparent throughout the spectra. The collection of lines above 5 Hz are due to mechanical resonances of the apparatus and optics while the broader structures between 0.1-5 Hz are due to seismic motion. Particularly, the broad structure between 0.2-0.3 Hz is due to the oceanic microseism. We believe that this is true torsional seismic motion and not due to residual cross-couplings as it is independent of small changes in either $\alpha$ or $\phi$. However, a second, co-located torsion balance is required to verify that the observed feature is truly the torsional microseism.

\section{Possible Applications}

The MINT apparatus, or similar instruments, have a wide range of applications. Here we will explore a collection of promising avenues.

\subsection{Inertial Seismic Sensing}

Seismic waves cause both translational and rotational motion. \cite{rotSeis} Traditionally, seismology has been limited to study only translational motion as rotation sensors did not meet the required sensitivity. Recently, a number of sensors have been developed to sense such motion \cite{ringGyro, BRS}. These observations have allowed a number of unique methods and studies. \cite{tiltseis, seisRing, rot} 

The MINT apparatus senses the torsional ground motion by using the pendulum as an inertial reference while having the optics rigidly attached to the ground. The sensed angle is then related to the rotation of the ground via:

\begin{equation}
\tilde{\theta}(\omega)=-\frac{\omega^2  }{\omega^2-\frac{i}{Q}\omega_0^2-\omega_0^2}\tilde{\theta}_g(\omega)
\end{equation}
where $\tilde{\theta}$ is the observed angular motion, $\tilde{\theta}_g$ is the angular motion of the ground, $\omega$ is the angular frequency of motion, $Q$ is the quality factor, and $\omega_0$ is the resonant angular frequency of the pendulum.

\begin {figure}[!h]
\centering
\includegraphics[width=0.5\textwidth]{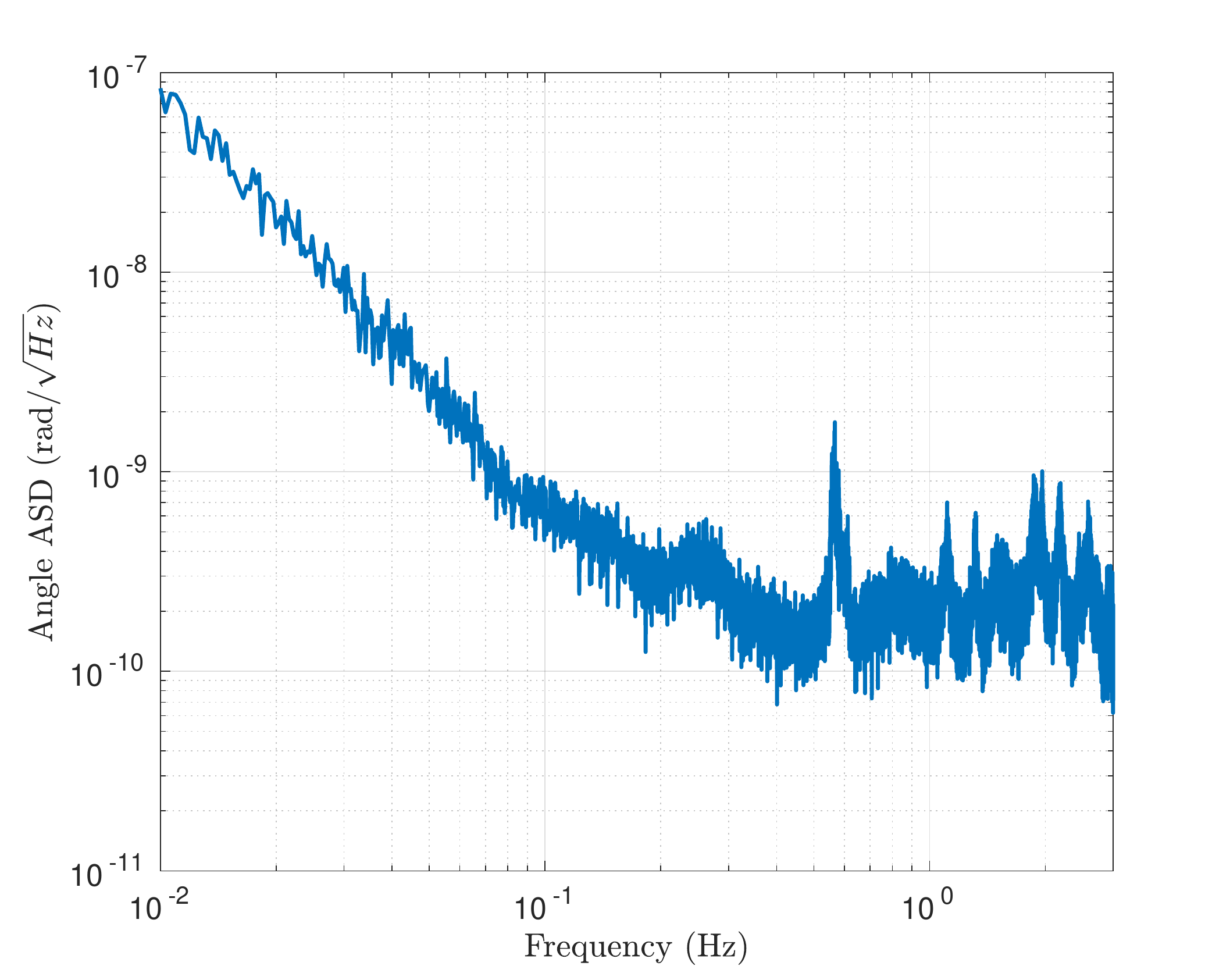}
\caption{Amplitude spectral density of the inertial angular noise.}
\label{ang}
\end{figure}

The inertial angular performance of MINT is shown in Figure~\ref{ang} which displays sub-$\text{nrad}/\sqrt{\text{Hz}}$ inertial angle noise down to 100 mHz. This performance allows for the observation of both teleseismic and regional Love waves.

\subsection{Gravitational Wave Observation}

If a $\times$-polarization gravitational wave passes vertically through the instrument, tidal forces cause an angular deflection of the pendulum which follows\cite{toba}:
\begin{equation}
\tilde{\theta}(\omega)=\frac{q^{12}}{2I}\frac{\omega^2 }{(\omega^2-\frac{i}{Q}\omega_0^2-\omega_0^2)}\tilde{h}_\times (\omega)
\end{equation}
where $q^{12}$ is the dynamic quadrupole, $I$ is the moment of inertia, and $h_\times$ is the gravitational wave strain. For the MINT pendulum the ratio $q^{12}/I\approx0.94$. Note that since this ratio is approximately one, the angular sensitivity shown in Figure \ref{ang} can be converted to strain by multiplying by a factor of two.

Although detection of gravitational waves at this sensitivity is not realistic and other instruments are much more sensitive to gravitational waves\cite{gracefo}, this apparatus can be used as a prototype for future torsion-balance based gravitational wave detectors.\cite{toba} Additionally, this instrument may allow for the study of atmospheric gravity gradient noise\cite{newtnoise} with the addition of an identical orthogonally-oriented apparatus.

\subsection{Elasto-gravity Signals}

The prompt gravitational signal caused by earthquakes has become a promising avenue for earthquake early warning. \cite{elasto} Torsion balances have begun to be explored as a method to observe these signals with high fidelity. \cite{torpedo} At its current sensitivity, MINT is expected to be capable of observing the elastogravity signals only from nearby earthquakes where the warning advance is minimal. However, such observations could increase the precision of earthquake source modeling. \cite{elastoModel} Additionally, with future upgrades we hope to reach observable distances with meaningful advanced warning.

\subsection{Torque Sensing}

\begin {figure}[!h]
\centering
\includegraphics[width=0.5\textwidth]{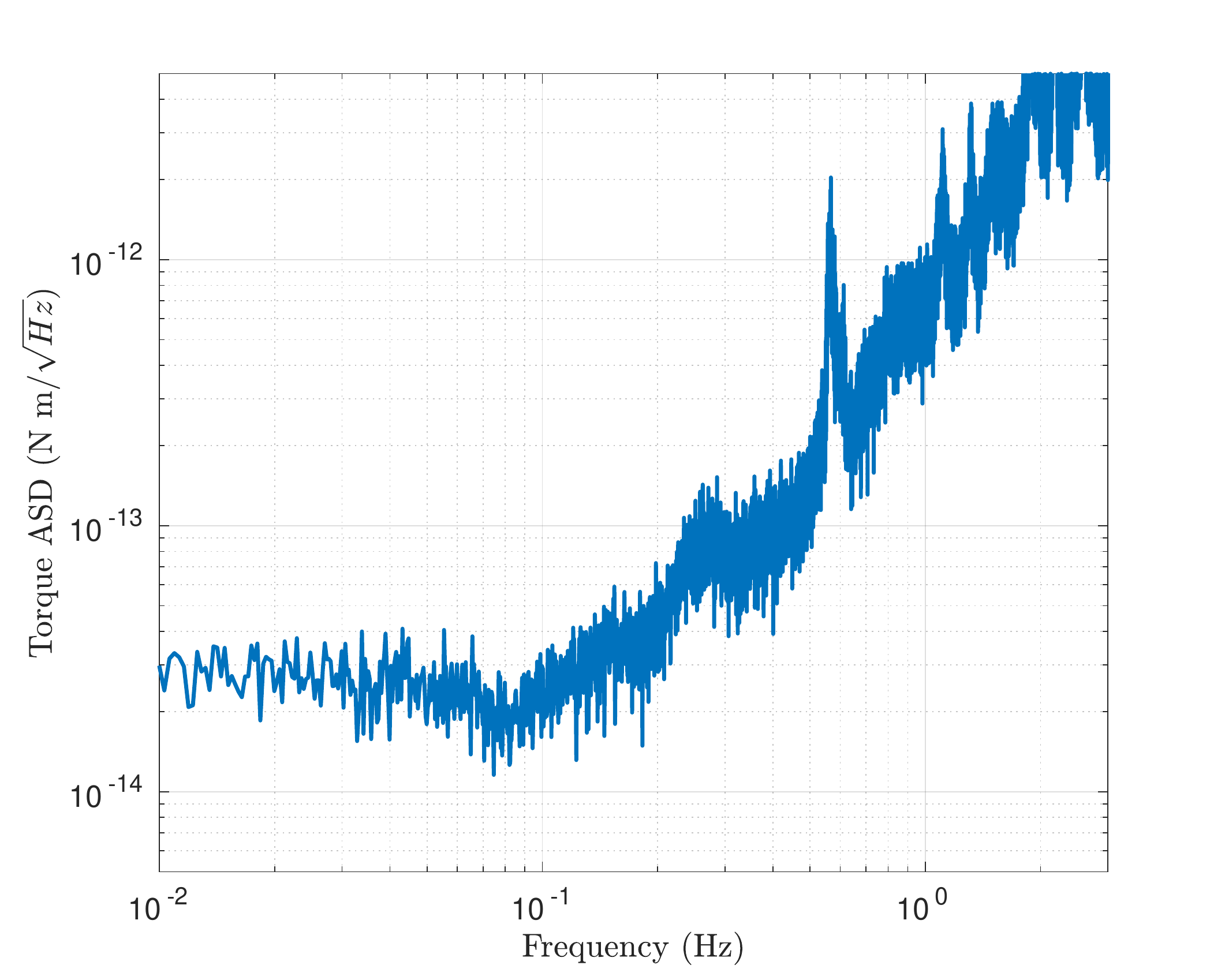}
\caption{Amplitude spectral density of the torque noise.}
\label{torq}
\end{figure}

Torsion balances are used to sense weak torques in a variety of experiments. These include short range gravity experiments\cite{shortRange}, tests of the equivalence principle\cite{EP}, and measurements of the gravitational constant\cite{bigG}. The torque sensitivity of the MINT apparatus is shown in Figure~\ref{torq}. The torque noise performance makes this instrument particularly promising for searches of ultra-light dark matter\cite{uldm}.

\section{Conclusion}

We have described a simple torsion pendulum with resonant frequency of 2.8 mHz whose angle is readout with both an autocollimator and a Michelson interferometer. The pendulum is locked in feedback using a pair of capacitive actuators in order to control the natural motion and to linearize the interferometric readout. This system achieves angular noise performance of $\sim200\ \text{prad}/\sqrt{\text{Hz}}$ between 0.2-30 Hz and $\sim10\ \text{prad}/\sqrt{\text{Hz}}$ above 100 Hz.

This apparatus can contribute to a variety of fields including rotational seismology, gravitational wave observation, and the study of elastogravity signals. Although first conceived of in the 18th-century, modernized torsion balances are at the forefront of today's precision measurement and provide the ideal apparatus for a number of scientific experiments.

\begin{acknowledgements}
This work was supported by funding from the NSF under Awards PHY-1607385, PHY-1607391, PHY-1912380 and PHY-1912514.
\end{acknowledgements}

\section*{Data Availability Statement}

The data that support the findings of this study are available from the corresponding author upon reasonable request.
\section*{References}
\bibliographystyle{unsrtnat}
\bibliography{MINT.bib}

\end{document}